# Theoretical and experimental study of AC loss in HTS single pancake coils

J Šouc, E Pardo, M Vojenčiak and F Gömöry
Institute of Electrical Engineering, Centre of Excellence CENG, Slovak Academy of Sciences, 841 04 Bratislava, Slovakia

e-mails: enric.pardo@savba.sk

**Abstract**

The electromagnetic properties of a pancake coil in AC regime as a function of the number of turns is studied theoretically and experimentally. Specifically, the AC loss, the coil critical current and the voltage signal are discussed. The coils are made of $Bi_2Sr_2Ca_2Cu_3O_{10}$/Ag (BiSCCO) tape, although the main qualitative results are also applicable to other kinds of superconducting tapes, such as coated conductors. The AC loss and the voltage signal are electrically measured using different pick up coils with the help of a transformer. One of them avoids dealing with the huge coil inductance. Besides, the critical current of the coils is experimentally determined by conventional DC measurements. Furthermore, the critical current, the AC loss and the voltage signal are simulated, showing a good agreement with the experiments. For all simulations, the field dependent critical current density inferred from DC measurements on a short tape sample is taken into account.

## 1 Introduction

One of the limiting factors for AC applications of the AC devices utilizing high temperature superconductors (HTS) is the electric power efficiency closely linked with the AC loss [1,2,3]. Indeed, in order to remove the heat produced by power dissipation in the superconductor it is required several times that power in a cryogenic system. For this reason, the AC loss in superconductors is a wide subject of study [4].

Many AC applications imply the use of windings, such as transformers and AC magnets. The current technologies produce HTS wires in the shape of tapes. Then, mainly due to mechanical reasons, one of the most feasible ways of manufacturing a winding is to form a pancake coil (or sets of piled pancake coils). Up to this time, only few theoretical works supported by experiments were dedicated to the AC loss of pancake coils [3,5,6,7,8,9,10].

From these references, in [5,8] the experimental results are compared to existing approximated analytical expressions or only a qualitative theoretical discussion is given. References [6,7] are based on relating the measured (or simulated) AC loss on a single tape under uniform applied field and the local average magnetic field of the coil. However, as discussed in [11], the interaction between tapes strongly influences the AC loss under a certain background external field. Actually, this can only be done when the field external to the tape is much larger than the tape self-field[1] and, therefore, the interaction between the turns is negligible [11,22]. Moreover, the assumption of uniform magnetic field cannot be done for a pancake coil, with a strongly non uniform field in the tapes [10,24]. Nevertheless, the measured AC loss from one tape can be directly used for predicting a coil AC loss for large coils made by piling many pancakes because the largest AC loss is produced in the top and bottom ends, where the radial field is huge and roughly uniform in each turn cross-section. Two other different approaches are done in references [3,9] but in both it is assumed that the local coil magnetic field is proportional to the current, which is considered uniformly distributed in the tape. Although these assumptions can be used at the critical current, they are not generally true [24]. Actually, there are very few works that rigorously calculate the AC loss in a pancake coil [10,24,12] and only in [10] it is compared with experiments. Moreover, a systematic study of the dependence on the number of turns in pancake coils has not been done.

[1] In fact, this condition is not strictly sufficient. The external field must be much larger than the self field of a stack made of as many tapes as those in the radial direction when the radial field dominates the AC loss (and equivalent with the axial direction when the dominant is the axial field).

Furthermore, in addition to the AC loss, the shape of the voltage signal in the coil is also interesting for both the understanding of the loss mechanisms and applications [13,14,15,16,17,18,19].

In this work, a detailed numerical and experimental study of the AC loss and the voltage signal in a single pancake coil as a function of the number of turns and the current amplitude is presented. The studied pancake coils are made of silver stabilized $Bi_2Sr_2Ca_2Cu_3O_{10}$ (BiSCCO) tape, although the main qualitative results are also applicable to coated conductors with a non magnetic substrate. Therefore, this work settles the basis for the future measurement and simulation of the AC loss in coated conductor coils for transformers or field generation applications.

This article is structured as follows. In section 2, the details of the studied pancake coils and its preparation are described. Following it, in section 3, we present the modelling methodology and some of their main results. Specifically, in section 3.1 the prediction of the coil critical current is detailed, as well as the extraction of the critical current density $J_c$ as a function of the magnetic field **B** from short tape samples measurements. Next, in section 3.2 the AC simulation method is presented, which takes into account the previously extracted $J_c(\mathbf{B})$. The measurement techniques are described in section 4. In the following section, 5, the AC loss experiments are compared with the measurements and they are discussed. Finally, in section 6 we present our conclusions.

## 2 Coil preparation

Six pancake coils with a different number of turns $N$ have been prepared (1, 3, 5, 10, 19 and 32 turns). BiSCCO tape[2] with cross section 3.8 mm x 0.2 mm and critical current $I_c$ = 50 A in self field have been utilized for the pancake construction. The inner diameter of all the pancakes is 17.1 cm. One side of the BiSCCO tape of total length 18.1 m have been covered by 0.08 mm Kapton insulating tape before winding the pancake. This insulating tape also the distance between the turns. Afterwards, the 32 turns pancake coil has been wound on a fibreglass cylinder with diameter 17.1 cm using such composed tape. The turns of the pancake have been fixed by Teflon tape on eight points symmetrically distributed on a perimeter of the pancake coil to suppress the mechanical vibration.

After performing all the experimental tests on the pancake coil with 32 turns (DC and AC experiments), 13 outer turns have been removed to obtain the pancake coil with 19 turns. Again, all the tests necessary for full DC and AC characterization have been done on this coil and subsequently 9 turns removed to produce the 10-turn pancake. The same procedure is repeated also for the rest of the set of pancake coils.

## 3 Modelling

### 3.1 *Pancake coil critical current, $I_{cp}$, simulation*

[2] Trithor superconductor

We have dedicated part of our effort to the development of a finite element procedure that could reliably predict the maximum current that the pancake is able to carry. The simulation have been performed using FemLab code and compared with measured data.

The critical current density $J_c$ of the BiSCCO superconducting tape depends on the local DC magnetic field and its orientation angle $\alpha$ with respect to the sample. This DC field can be divided into two components: parallel $B_{||}$ ($\alpha$ = 0 deg) and perpendicular $B_{\perp}$ ($\alpha$ = 90 deg) with respect to the wide face of the tape. Due to intrinsic properties of the BiSCCO material and producing tape technology the influence of the perpendicular component $B_{\perp}$ on $J_c$ is large. As a consequence the anisotropic dependence of the local critical current density on magnetic field can be described by the semi-empirical formula

$$J_c(B_{||}, B_{\perp}) = \frac{J_{c0}}{\left(1+\frac{\sqrt{k^2 \cdot B_{||}^2 + B_{\perp}^2}}{B_0}\right)^{\beta}} \qquad (1)$$

where $J_{c0}$, $k$, $B_0$, $\beta$ are constants characterizing the HTS material [20]. This formula was used for calculation of the local critical current density $J_c$ in the whole cross-section of the conductor.

As a first step for simulating of the pancake critical current $I_{cp}$ the critical current $I_{c,tape}$ of the short sample of length 10 cm exposed to the external magnetic field with various orientations is investigated. The constants $J_{c0}$, $k$, $B_0$, $\beta$ are estimated by comparing the calculated critical current, obtained by integrating of the $J_{cloc}$ over the superconductor cross-section $S$:

$$I_{c,tape} = \int_{S_{tape}} J_c(x, y)dS \qquad (2)$$

with experimental data. A good agreement between measurement and calculation is found, figure 1, when using the parameters $J_{c0} = 1.34\times10^8$ A/m$^2$ , $k$ = 0.1, $B_0$ = 0.008 T, $\beta$ = 0.58.

After determining in this way the $J_c(B_{||}, B_{\perp})$ dependence for the HTS tape, a similar procedure is applied to the calculation of the critical current $I_{c,turn}$ of the individual pancake turns. This value can be generally different for each turn because individual turns are exposed to the magnetic field generated by other turns. Here, we present the results of calculation made separately for each turn, considering the same current uniformly distributed in the other turn cross-sections, except the turn for which $I_c$ is calculated. From these calculations the critical currents of all individual turns are determined in dependence on the turn number starting from the inner part of the pancake - figure 2. The critical current is higher at the beginning and at the end of the pancake winding. This is because the magnetic field in these parts is oriented mainly parallel with respect to the tape.

Once the $I_{c,turn}$ values in individual turns are known, the voltage on every turn is calculated assuming the power law dependence of the electric field intensity $E_t$

$$E_t = E_c\left(\frac{I}{I_{c,turn}}\right)^n \qquad (3)$$

taking $E_c$ = 1μV/cm as the criteria for critical current. The $n$ value used in this estimation is that one obtained on the short sample in conditions when $I_{c,tape}$ reaches the same value as $I_{c,turn}$. The total voltage $U_{tot}$ on the pancake coil is calculated by summing the voltages on individual turns. $I_{cp}$ and $n$ value of the coil are then calculated as fitting parameters of the $U_{tot}(I)$ dependence. The comparison of the simulated and measured $I_{cp}$ is presented in section 5.

### 3.2 *Pancake coil AC loss simulation*

It has been shown that the electromagnetic properties of a superconductor can be predicted by means of the minimum magnetic energy variation (MMEV) method [21,22,23], which assumes the critical state model. Recently, this method has been applied to pancake coils with a constant critical current density $J_c$, [24]. However, as shown in the previous section, the response of the measured BiSCCO pancake coil is significantly influenced by the field dependence of $J_c$. In this article, the numerical model in [24] is extended in order to take into account the field and position dependence of $J_c$.

### *3.2.1 Procedure for the current distribution.*

In order to calculate the AC loss and the voltage in the coil, it is necessary to solve first the current distribution in the superconductor. In this section, the numerical model presented in [24] for constant $J_c$ is extended for an arbitrary magnetic field and position dependence. The method is based on minimum magnetic energy variation (MMEV), which assumes the critical state model.

The physical grounds of the MMEV are the following (more details can be found in [22,24]). First, we assume that the pancake coil has cylindrical symmetry, something that can always be done for closely packed coils. Then, in the Coulomb's gauge ($\nabla \cdot \mathbf{A}=0$), both the current distribution **J** and the vector potential **A** follow the angular direction and can be treated as scalars, $J$ and $A$. For the critical state model, $J$ minimizes a certain functional $F$ [26]. Following the same formalism as [22], the functional is

$$F[J(t)] = \frac{1}{2}\int_\Omega [J(t) - J(t-\Delta t)][A_J(t) - A_J(t-\Delta t)]\mathrm{d}\Omega \qquad (4)$$

where $A_J(t)$ is the vector potential created by $J$ at a certain time $t$, $\Delta t$ is a very small positive time interval and $\Omega$ is the superconductor volume. The functional (4) has to be minimized with two constrains: $J$ cannot be larger than the critical current density, $J_c$, and the total current flowing in the superconductor is fixed. Then, the $J$ change in time is originated by a change in $I$.

In order to solve $J$ by minimizing $F$ in (4), we make the following discretization of the superconducting volume and $J$ values. Each turn of the coil is divided into ring elements with identical rectangular cross-section, as follows. A rectangular area that contains the turn cross-section is taken, uniformly divided and the elements outside the turn cross-section are considered free of any current. For simplicity, we assume an elliptical tape cross-section. Similarly as in [22,24], we only allow a certain discrete number $n_J$ of nonzero values of $|J|$, in such a way that the possible values of $J$ are equally spaced. Since the maximum possible $|J|$ is $J_c$ at zero field, $J_{c0}$, we take $J=i_J\, J_{c0}/n_{J0}$, where $i_J$ can be any integer between the given values $n_{J0}$ and - $n_{J0}$. A non uniform $J_c$ can be simulated by setting a maximum allowed $|J|/J_{c0}=n_J$ as a function of the element position. A $J_c(\mathbf{B})$ dependence can be implemented in the same way using $n_J=\mathrm{int}[n_{J0}J_c(\mathbf{B})/J_{c0}+1/2]$.

Next, we describe the procedure for the $F$ minimization. For a constant $J_c$, the minimization method for a given change in $I$ is the same as in [22] with the mutual inductances in [24]. In the present article, however, the algorithm has to be changed. Since $J_c$ and $J$ depend on the magnetic field and the magnetic field depends on $J$, the problem has to be solved iteratively. First, we describe the case of a single turn coil. We start with the zero-field cool situation with no transport current, that is $J$=0 and **B**=0 everywhere. After increasing the current up to a certain amount, we calculate the current distribution $J$, by numerically minimizing $F$ in he same way as in [22], assuming a constant $J_c$ with value $J_{c0}$. With this, we finish the first iteration. The next one starts by computing **B**, from which $J_c$ in each element is calculated according to the $J_c(\mathbf{B})$ relation (1). Then, $|J|$ in the elements where it exceeds the new $J_c$ is "cut" to $J_c$. This changes $A_J$ in (4) and, therefore, it is recalculated. In addition, the "cut" in |J| may modify the total current by a certain amount $\Delta I$. For this reason, we afterwards add (or remove) a current $-\Delta I$ distributed in the way that minimizes $F$, using the recently computed $J_c$. The iterations are repeated until the $J$ change in any element after iteration is no more than the interval in the discretized current density, $J_{c0}/n_{J0}$.

After this condition is achieved, we impose 5 more iterations in order to ensure a stable result. Actually, we found that the results are insensitive to the number of extra iterations for a large enough $n_{J0}$. The current distribution for a subsequent current is obtained in the same way but starting with **B** for the previous current. Following this procedure, the whole AC loop can be calculated. The case for a coil with many turns is solved in the same way but taking into account that the current variation due to "cutting" |J| to $J_c$ after computing the magnetic field is different for every turn.

### *3.2.2 Critical current.*

Using the method above, it is also possible to predict a certain coil critical current $I_{cp}$, defined as the current for which $J$ in some turn of the coil equals to the local $J_c$ for all the cross-section[3]. This $I_{cp}$ is essentially the same as that for the turn with minimum critical current from the procedure in section 3.1. Although the minimum $I_c$ is not always the same as that for the whole coil[4], in figure 2 it is seen that for a large enough number of turns, $I_c$ is roughly uniform in the whole pancake coil and, therefore, the minimum $I_c$ is a good approximation. In any case, the MMEV calculations yield the correct $I_c$ for any single-turn coil. Since the limit of a very large radius corresponds to the infinitely long geometry, this method can be used for finding $I_c$ for any tape cross-section from a known $J_c(\mathbf{B})$.

### *3.2.3 Main features of the current density and the magnetic field.*

An example of the calculated current distribution and magnetic field lines is presented in figure 3, corresponding to a 19 turns coil at the peak of the AC current, with amplitude 14.1 A and $I_{cp}$ calculated by MMEV of 29.4 A. As expected [24], there appears a zone with null average current across the tape thickness (where $J$ is either null or there is the same amount of positive and negative $J$) with approximately the same width for all the turns. The net current is transported in the remaining top and bottom regions with approximately the same distribution for all the turns. The magnitude of $J$ is, consistently, $J_{c0}$ next to the neutral zone (place where $J$=0), since there **B**=0, and $J$ increases when moving away from the neutral zone. At the top and bottom zones, $|J|$ is significantly lower than $J_{c0}$ due to a considerable $B_\perp$, while $|J|$ is large in the remaining cross-section, where the radial field practically vanishes.

A detail of the radial field distribution is plot in figure 4. It is seen that, except from the two more external turns, the radial field is roughly the same for all of them.

Figures 3 and 4 suggest that the behaviour of a coil with low separation between the turns is approximately the same as a slab with a $z$ dependent $J_c$, proportional to the fraction of superconducting material in every $z$ position. Indeed, for this case, the transport current has to distribute like a slab because it is the only way to create a zero radial field in the superconducting region [24]. Besides, this configuration creates a radial field roughly uniform. The current density around the centre of the tapes, with positive and negative values, shields the vertical field but does not significantly contribute to the radial field.

For a transport current lower than the ac peak value, there are more complicated effects on the $J$ magnitude. This can be seen in figure 5, where the current penetration process of the 19-turn coil is represented by $J$ as a function of $z$ at the central $r$ of the 11th turn. This turn is chosen for simplicity because it presents a current-free core. In figure 5, it is seen that for positive current there appears a peak in $J$, corresponding to the place where $B_r$ vanishes.

---

3 Actually, the MMEV method assumes the critical state model and it is not possible to take into account the overcritical situation.

4 This affirmation applies when taking into account that the actual superconductor follows a certain smooth $E(J)$ relation and Ic is defined as the current that creates a certain threshold average electrical field, $E_c$ (see section 3.1). For a hypothetical sharp $E(J)$ relation at $J=J_c$, the pancake critical current corresponds to the minimum one.

### 3.2.4 AC loss and voltage

Once the current distribution for the whole AC cycle is obtained, the voltage drop in the coil and the AC loss are calculated following [24], by means of (3) and (4) in that reference. In essence, the AC loss is calculated from the volume and time integration of **J**·**E**, where **E** is the electrical field. The later, as well as the voltage drop, are obtained from the vector potential which is computed from the current distribution by the Biot and Savart law.

## 3.3 On the voltage compensation

The properties of the superconductor can be qualitatively studied from the voltage signal [13,14,15,16,17,18,19]. However, the voltage in a coil where it circulates an AC current $I=I_m\cos\omega t$ has a very large inductive component (proportional to $\sin\omega t$) which does not depend on the material properties, neither contribute to the AC loss. Therefore, such inductive signal should be appropriately removed in order to make a detailed analysis. We mention the removal of this component as compensation. One criterion is to remove a signal proportional to $\sin\omega t$ in such a way that the remaining voltage is only due to the flux penetration in the superconductor. Actually, for low current (or low applied field) the superconductor experiences perfect shielding, with no flux (or current) in the volume, and the response is linear and lossless. Therefore, the voltage for the low current limit can be expressed as $V=\omega L_0 I_m\sin\omega t$, where the constant $L_0$ is the low field self inductance. Once $L_0$ is known, the compensated voltage $V_c=V-\omega L_0 I_m\sin\omega t$, where $V$ is the total voltage, is purely due to the flux penetration in the superconductor. The value of $L_0$ can be obtained as the low $I_m$ limit of the first imaginary harmonic[5]. From the experimental point of view, the inductive component can be either removed by means of a variable mutual inductance coupled to the transport current or by properly amplifying the signal of the Rogowski coil in figure 6, as done in this article. Then, the gain of the amplifier is set in order to cancel the inductive part at low current[6] [25].

# 4 Experimental method

For the experimental tests, the pancake coil is immersed in liquid nitrogen bath. The critical current $I_{cp}$ is determined for every pancake by monitoring the voltage measured on the whole pancake during increasing of the DC current. For the determination of $I_{cp}$, the criterion 1 μV/cm is chosen.

The AC experiments are done in a transformer configuration. As experimentally observed in section 4.4, the losses in the coil are not influenced by the presence of the transformer. A schematic view of the arrangement is shown in figure 6a and its photo is displayed in figure 6b. The coil is the secondary of the transformer and its beginning and end are connected to each other by means of a copper bridge, making a short circuit. In this configuration, an AC sinusoidal current in the primary creates a sinusoidal current in the secondary. The measurements were done at the frequencies $f$ = 36 Hz, 72 Hz and 144 Hz. In order to measure the current in the secondary, a Robowski coil is placed in the coil connexion, made of transposed copper braid in order to minimize the eddy currents. The superconducting coil is kept at liquid nitrogen temperature by means of a non-metallic cryostat. An advantage of this arrangement is that no current leads enter the bath of cryogenic liquid. Moreover, it reproduces the case of a superconducting transformer.

---

5 Alternatively, $L_0$ can be found as that which $V_c$ at low amplitudes has a real and imaginary harmonics with the same value, as it is predicted for a circular wire [15]. Actually, both ways of determinig $L_0$ are equivalent, since for the low field limit the first real harmonic approach to zero.

6 (see previous footnote)

To determine the AC loss, it is enough to register the component of the first harmonic of the voltage that is in phase with the AC sinusoidal current. Multiplying the rms values of these two quantities gives the power dissipated in the coil[7]. The voltage in the coil (or a segment of it) can be measured by the following two methods.

### *4.1 Method A (representative turn method)*

In this method, instead of the whole coil voltage, it is measured only the signal from the pair of taps placed on one turn, making a C shaped loop (figure 7). As discussed in [10], if the voltage felt by these taps, $U_A$, is independent on the turn where it is placed, it is representative of the loss per unit length. Indeed, in figure 4 (and [24]) we showed that the radial field, which is responsible of the flux derivative that produces the measured voltage, is roughly uniform in the turns close to the average radius. However, it is significantly smaller in the inner and outer turns. Then, opposite than in [10] where the loop is located in the outer turn, we place the measuring taps in a turn close to the mid coil radius, resulting in a much higher performance (section 4.4). The advantage of using these voltage taps is that the huge coil inductive voltage is avoided, although there is still an inductive signal (but much smaller) that has to be compensated.

Although placing the C loop in any turn close to the mid radius gives mainly the same results, in our case, the proper turn is chosen on the base of the AC simulation of the current distribution – (section 3.2, figure 4). Specifically, we choose the turn which the current is distributed symmetrically with a neutral zone, where $J$=0, at its centre. For the pancake with 32, 19, 10, 5 and 3 turns this corresponds to the turn number 19, 11, 6, 3 and 2 counted from the pancake inner side. The total loss of the coil is then $P_{coil} = \mathrm{Re}\{U_{A,1,rms}\}\times I_{rms}\times L/l$, where the subindex "rms" indicates rms values, $\mathrm{Re}\{U_{A,1}\}$ is the part of the first harmonic of the voltage which is in phase with the secondary current $I$, $l$ is the distance between taps and $L$ is the total length of BiSCCO tape in the pancake. In our experiments, the distance between taps $l$ is 3 cm. To compare the properties of coils with different number of turns and experiments at different frequencies of AC current, the loss per cycle and unit length of tape is evaluated as $Q_{tape} = \mathrm{Re}\{U_{A,1,rms}\}\times I_{rms}/(l\,f)$.

### *4.2 Method B (secondary voltage method)*

This method is applicable only in the case of supplying the AC current via the transformer technique. Its principle and detailed description has been published in our previous work [25]. It uses a pick-up loop embracing the core with the same number of turns as the measured coil has (figure 7). In this way, the voltage $U_B$ induced in this pick-up coil is identical to the voltage on the whole secondary circuit, including the Cu braid short-circuiting the coil ends. This loop must be close to the coil position in order to get the same flux from the transformer as the coil (usually, the flux in a transformer core is not the same in all its length). The total power dissipated in the secondary can be determined as $\mathrm{Re}\{U_{B,rms}\} \times I_{rms}$, where $\mathrm{Re}\{U_B\}$ is the part of the whole secondary voltage which is in phase with the secondary AC current $I$. To eliminate the contribution due to dissipation in Cu braid the difference $U_B - U_R$ ($U_R$ is the voltage measured on Cu bridge) instead of $U_B$ is taken in the measurements for all the considered pancakes. In contrast to the method A, where by measuring of $U_A$ only the pancake segment corresponding to the distance between taps $l$ is characterized, using the method B the properties of the whole pancake are measured. Naturally, in the case of a homogeneous pancake the results of both methods should be the same.

---

[7] In general, the loss in one cycle is the current times the voltage integrated in one period. For a sinusoidal current, only the in phase fundamental harmonic of the voltage contributes to the loss because the rest of the harmonics vanish in one cycle integration.

### *4.3 AC loss*

The in-phase component of the voltage signals can be measured by a phase-sensitive voltmeter, such as a lock-in amplifier. The AC current in the coil is measured by a Rogowski coil. Its signal is also utilized to compensate a substantial portion of the out-of-phase component of the measured voltages. This compensation is more important in the secondary voltage method, in which a huge inductive component appeared (section 3.3) [25].

### *4.4 Method checks and comparison*

In this section, we check that the AC loss measured from methods A and B is correct. This is done by comparing with the conventional technique. In that case, the coil is removed from the transformer and the voltage is measured by voltage taps at the coil connections to the current terminals. In this way, a possible interaction with the transformer ferromagnetic core is avoided.

In Figure 8 there are plot the 32-turns-pancake losses at 36 Hz determined by the representative turn technique (choosing the correct turn and the outer turn, respectively, for voltage taps placing) compared with the results obtained by the secondary voltage method and the conventional way. As can be seen, all measurements yield the same results, within the measurement dispersion, except those from Method A when the outer turn is used for placing the voltage taps. This comparison indicates the following:

- The AC loss of the coil forming transformer secondary winding can be measured without (measurable) AC loss contribution of the ferromagnetic core itself. This is important for method B which requires a transformer. Of course, the representative turn method A is usable for pancake coil placed as a secondary or placed outside of the transformer.
- It is important to choose the proper turn for placing the voltage taps when Method A is used. In other case, as it was shown by simulation of the magnetic field distribution (section 3.2), not all the relevant flux is felt and a huge error in the reduced voltage signal (up to one order of magnitude) can appear.
- From method A only the AC loss of a pancake segment is detected, while method B (and the conventional one) measures the loss in the whole coil. Then, the identity of the results confirms the homogeneity of the pancake.

In addition to being closer to an application, the transformer set-up allows to easily apply large AC currents to the coil, in contrast to the conventional technique. Methods A and B can be used equivalently for one homogenous pancake. However, using of the method A is not straightforward for more complicated system comprised of several pancakes because the correct position of the voltage taps if unknown. Method B, which is more general, should be used in such case.

In figure 9 the AC loss per cycle measured on the 10-turn pancake at frequency $f$ = 36 Hz, 72 Hz and 144 Hz by method A (representative turn technique) and method B (secondary voltage method) are plotted, showing a perfect agreement. Methods A and B have also been compared for all the other pancakes, obtaining a poorer agreement only for 3 and 5 turns, for which the assumptions done for method A are no longer fulfilled (section 4.1). Therefore, in the rest of the measurements we use either method A and B for one single turn or 10 or more turns but only method B for 3 and 5 turns.

## 5 Results and discussion

## 5.1 Critical current

$I_c$ measurements of short samples taken from the tape used for the pancakes manufacturing reveals a critical current scatter around 10 percent. The minimum and maximum measured values are 44 A and 50 A respectively. Therefore the simulation of the pancakes critical current have been performed for two cases assuming the short sample critical current 46 A [with the parameters in (1): $J_{c0} = 1.54\times10^8$ A/m$^2$ , $k$ = 0.12, $B_0$ = 0.0038 T, $\beta$ = 0.47] and 50 A ($J_{c0}$ = $1.34\times10^8$ A/m$^2$ , $k$ = 0.1, $B_0$ = 0.008 T, $\beta$ = 0.58), respectively.

The calculated and measured critical currents of pancakes $I_{cp}$ with different number of turns are compared in figure 10. The critical current is lower for the pancakes with higher number of turns because of a larger magnetic field affecting the individual turns.

## 5.2 Ac loss

AC loss measurements at various frequencies are very useful to identify the prevailing loss mechanism. The data in figure 9 show that the loss per cycle (and unit length) is practically independent on the AC frequency. Consequently, the loss mechanism in the investigated coils is well described as the hysteretic loss. Therefore, in the following, only the data taken at the frequency of 72 Hz are presented.

The comparison of the AC loss behaviour of all the pancake coils is shown in figure 11. In this plot, the power dissipation per unit length of the tape used in the pancake is shown. The short sample (10 cm) self-field loss is shown in the picture as well. The results for 1 turn coil and short sample are identical in the frame of the measurement precision. Obviously, the diameter of the loop (17.1 cm) is large enough to produce a magnetic field not very different from the self-field of a straight tape. Besides, one can see that increasing the number of turns in the pancake increases the dissipation per unit length of the tape. Evidently, it is due to increasing of the magnetic field to which the pancake turns are exposed.

Figure 12 shows the comparison of the measured losses with theoretical predictions. The quantity shown is the loss factor $\Gamma = 2\pi Q/(\mu_0 I_m^2)$, where $I_m$ is the amplitude of the AC current and $Q$ is the loss per cycle and unit length. The dependences are plotted with respect to the normalized current $F = I_m/I_{cp}$. For the experimental curves, it is used the measured $I_{cp}$ for each pancake coil (figure 10). The $I_{cp}$ for the theoretical curves has been found by the simulations described in the section 3.1. Generally, very good agreement between simulated and measured pancakes AC loss data is found. This is a bit surprising when taking into account the simplifications used in the AC loss calculation. For example, the superconducting zone of the BiSCCO tape is replaced by a bar with elliptic cross-section, ignoring the non homogeneity effects originated by the actual filamentary structure [17].

The dependence on the number of turns $N$ of the ac loss is shown in figure 13 for a specific current ($I_{rms}$=19.4A). For low $N$, the AC loss increases roughly linearly (also in linear scale), consistent with the roughly linear increase of the radial field when superposing several turns. For larger $N$, the computed AC loss at fixed current amplitude (practically) saturates for a large number of turns (roughly 50). This is because for a large $N$, the radial magnetic field is roughly uniform and the dependence of $J_c$ on the axial (parallel to the tape) field is very weak. However, the experiments show that, although the loss per unit length becomes lower than the extrapolated linear dependence for low $N$, it is far from being saturated. This may be caused by a larger sensitivity than expected of $J_c$ on the magnetic field.

## 5.3 Voltage signal.

To support the correctness of the measurement, the shape of the measured voltage signal for different frequencies and pancake transport currents has been registered and compared with the signals determined by the simulations, described in section 3.2. Moreover, as discussed here, the voltage signal gives extra qualitative information on the superconducting tape. The inductive part of all the presented signals is compensated (section 3.3).

In figure 14, the measured and calculated waveforms for different pancakes carrying a current of 20 A rms at $f$ = 72 Hz are shown. Besides, the waveforms for the 32 turns pancake at different AC current, $I_{rms}$=4.5 up to 30A, have been measured and are presented in figure 15. Good qualitative coincidence in the shape of the signals between experiment and theory is found, showing the following main features.

A minimum around zero transport current is clearly observed in the measurements and the simulations, figures 14 and 15. This kink is more pronounced for pancakes with higher turn number or higher current amplitudes and, thus, when the superconductor is exposed to higher local magnetic field amplitudes. The position of this minimum, slightly before $I$=0, approaches to $I$=0 with increasing the number of turns. The appearance of the kink in the voltage and its behaviour can be explained from the field dependence of $J_c$. Indeed, the voltage signal for a constant $J_c$ was also calculated, showing no minimum. The origin of the minimum is the following. The critical current density is dominated by the radial field. A coil with a large number of turns can be approximated to an slab (section 3.2.3). In particular, $B_r$ at the coil top and bottom surfaces is practically uniform and proportional to $I$. For $I$=0, $B_r$ vanishes at the surface and $J_c$ there is the maximum possible. This can be seen in figure 5. Using the slab approximation, the compensated voltage is the time derivative of the flux between the coil mid plane (z=0) and the top (or bottom) surface. Additionally, a current on the surface does not create flux inside a slab (or the coil). Therefore, if the induced current d$I$ during a differential time interval stays on the surface, the flux does not change and the voltage is zero. The closest situation to this one is when $I$=0 because $J_c$ at the surface is the maximum possible and d$I$ stays as close as possible to the surface. Then, the flux change due to d$I$ is minimal and also is the voltage. The depth and the position of the voltage minimum depend on the number of turns. If the number of turns is not very large, the slab approximation is less valid. Specifically, $B_r$ on the surface is no longer proportional to $I$ and the current distribution closer to the surface contributes more to $B_r$ there. Then, for the half cycle with increasing $I$, $B_r$ at $I$=0 is positive and $B_r$ is zero (or minimum) for a negative $I$. This explains why the measured minimum is before $I$=0, approaching to $I$=0 with increasing the number of turns. The kink in the voltage signal is more pronounced for larger number of turns and current amplitudes because the local $B_r$ amplitude is higher, causing a larger $J_c$ difference between the current peak and $I$=0.

The measured amplitude of the compensated voltage signal is lower than the AC loss simulations, which assume the critical state model (figure 14). An explanation of this effect was shown in [16] where the power low $E \approx I^n$ dependence with different $n$ was taken into account for voltage signal calculation in a tape carrying an AC transport current. As was shown in that work, with $n$ decreasing from a high value (critical state approximation) to a lower value (more real superconductor case) the absolute value of the voltage signal decreases.

For a sufficiently large AC current, there appears a peak in the voltage signal at the position where the AC current reaches its maximum and overcomes the pancake critical current. This is seen in figure 15 for 32 turns and $I_{rms}$=30A, although the same effect appeared for all the coils.

## 6 Summary and conclusions

In this article, we have presented a complete theoretical and experimental study of the AC response of a pancake coil as a function of the number of turns and the AC current.

A lot of effort has been put in the optimization of the AC loss measurement, which is done inductively with the help of a transformer. Additionally, thanks to an AC loss simulation, it has been found that the AC loss can be measured by means of voltage taps on a turn close to the coil average radius (with or without transformer), avoiding the difficulties in correcting the huge inductive signal of the whole coil. However, the later technique is only accurate for large number of turns (≥10). Both methods have been successfully checked by comparing with the measurements from conventional voltage taps at the current leads.

The AC loss computations, which take into account the field dependence of the local critical current density, have shown a very good agreement with the experiments. Thus, it is seen that the presented computing technique, based on the minimization of the magnetic energy variation (MMEV), is a powerful tool for the prediction of the AC loss in coils. These AC loss calculations were able to be done thanks to the extraction of the $J_c(\mathbf{B})$ dependence by means of DC measurements on short tape samples and another self-consistent model. Furthermore, by means the latter computations, the whole coil critical current has been predicted, showing again a good agreement with the experiments. It is worth to mention that the measured variation of the critical current in the length of the used tape was around 10%. Therefore, any prediction cannot be more accurate than this fraction.

We have observed that, as expected, the coil critical current $I_{cp}$ decreases with the number of turns $N$, although it practically saturates for large $N$. This can be explained as follows. The dependence of $J_c$ on the axial field is very weak so the dominating component is the radial field $B_r$. For a pancake coil, $B_r$ is roughly uniform and it saturates for a large number of turns [24]. Then, $I_{cp}$ must saturate for a large number of turns. Moreover, the dependence of $J_c$ on $B_\perp$ in (1) contributes to this effect because $J_c$ is less sensitive to $B_\perp$ variations at larger fields (see figure 1). For the same reasons, the computed AC loss per unit length $Q$ at a fixed current amplitude also (practically) saturates for a large number of turns (roughly 50). However, the experiments show that, although the loss per unit length becomes less sensitive at the largest number of turns, it is far from being saturated. This may be caused by a larger dependence than expected on the magnetic field in some segments of the tape.

In conclusion, we reported a full methodology for the measurement and prediction of the AC loss (and voltage signal) in generic superconducting pancake coils.

**Acknowledgements**

This work was supported by the Slovak Research and Development Agency under the contract No. APVV-51 – 045605, the European Atomic Energy Community under the contract FU07-CT-2007-00051 and the NESPA training network from the European Commission under contract number MRTN-CT-2006-035619.

Figure 1. Measured (open symbols) and simulated critical currents of the short BiSCCO sample in dependence on the applied DC magnetic field $B_{ap}$ with different orientation angles. The zero orientation is set to the direction parallel to the magnetic field.

Figure 2. Simulated $I_{c,\ tape}$ of individual turns as a function of the turn number counting from the interior for a pancake with 3, 5, 10, 19 and 32 turns (open symbols). For comparison, the measured $I_c$ for a short sample is also shown by a full point.

Figure 3. Computed supercurrent distribution, $J$, in the 19 turn coil cross-section at the AC current peak with amplitude $I_m$=14.1A and calculated critical current $I_{cp}$=29.4A (from section 3.2). The lines are the magnetic field lines calculated as $rA$ level curves, where $A$ is the vector potential and $r$ is the radial coordinate. This coordinate increases from left to right. The horizontal and vertical axis are in the same scale.

Figure 4. Calculated radial magnetic field, $B_r$, for the same situation as figure 3 as a function of the radial coordinate $r$ at several vertical positions $z$. The vertical origin $z$=0 has been chosen in the midplane and $w$ is the tape width.

Figure 5. Current density $J$ as a function of the vertical position $z$ at several currents at the centre of the 11$^{th}$ turn in the 19-turn coil for a current amplitude $I_m$=28.3 A (20 A rms). The curves are for the instantaneous current values $I/I_m$=-2/3,-1/3,0,1/3,2/3,1 in the arrow direction. The $z$ origin is at the centre of the coil.

Figure 6. (a) set-up of the experiment and (b) photo of the experimental apparatus. The inserted picture in (b) is the pancake wound on the fiberglass cylinder

Figure 7. Principle of the method A (measured $U_A$) and method B (secondary voltage method, measured $U_B$-$U_R$).

Figure 8. Comparison of the AC loss per cycle and unit legth $Q$ of the pancake coil with 32 turns measured by method A with AC loss of the same pancake extracted by the conventional taps at the current leads. For method A, we either placed the C-shaped loop, at the coil mid radius (correct loop), or on the outer loop. The frequency of the AC current was $f$ = 36 Hz.

Figure 9. Comparison AC loss per cycle and unite length for the 10 turns pancake measured by the methods A and B at several frequencies: 36, 72 and 144 Hz.

Figure 10. Comparison of the calculated and measured critical current $I_{cp}$ dependences for pancakes with different number of turns. The calculation was performed for a short sample with $I_c$ = 50 A and 46 A (see text).

Figure 11. Measured AC loss dependences for individual pancakes. The AC loss of a short sample is shown as well, $f$=72 Hz.

Figure 12. Comparison of the simulated and measured reduced pancakes AC losses represented by $\Gamma = 2\pi Q/(\mu_0 I_m^2)$ on the reduced transport current $F = I_m/I_{cp}$ .

Figure 13. AC loss per cycle and unit length as a function of the number of turns.

Figure 14. Thick lines are for the voltage signal waveforms monitored for individual pancakes at transport current 20 A rms, *f* = 72 Hz. Lines with symbols represent the simulated waveforms for individual pancakes at the same conditions; only half period is shown. The thin line is proportional to the current.

Figure 15. Waveforms of the voltage signals monitored for the 32 turns pancake at different transport current amplitudes, *f* = 72 Hz.

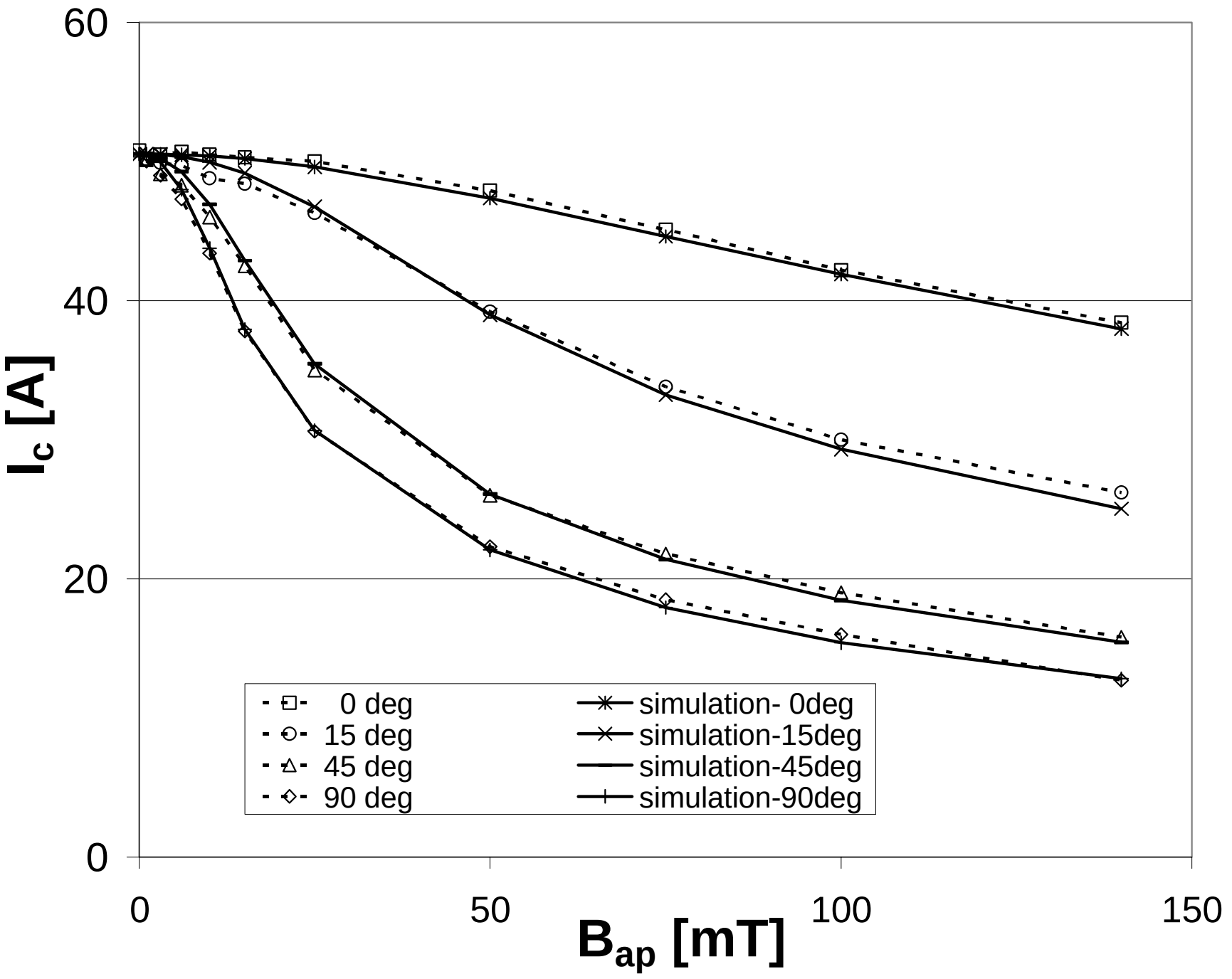

**Figure 1**

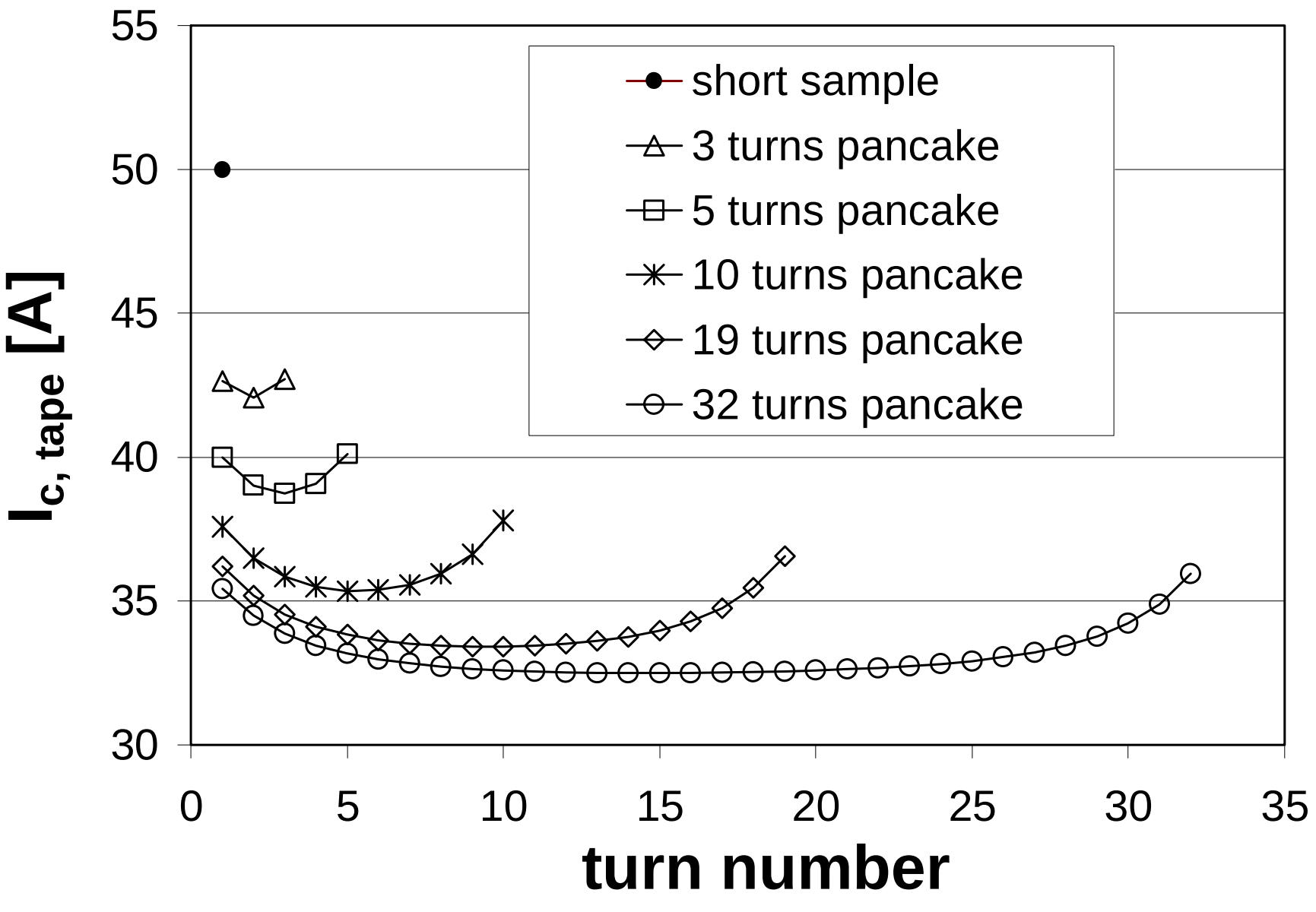

**Figure 2**

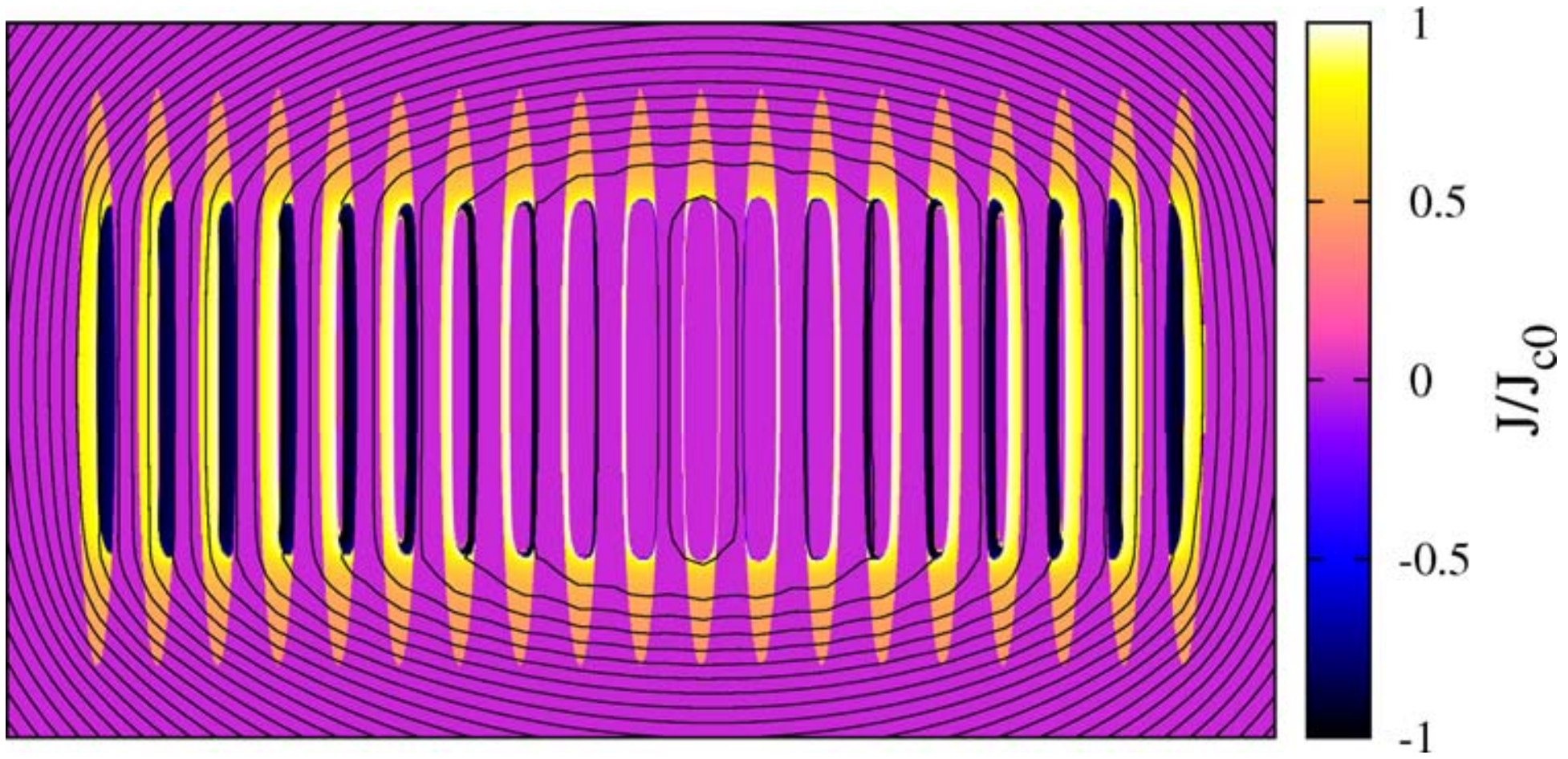

**Figure 3**

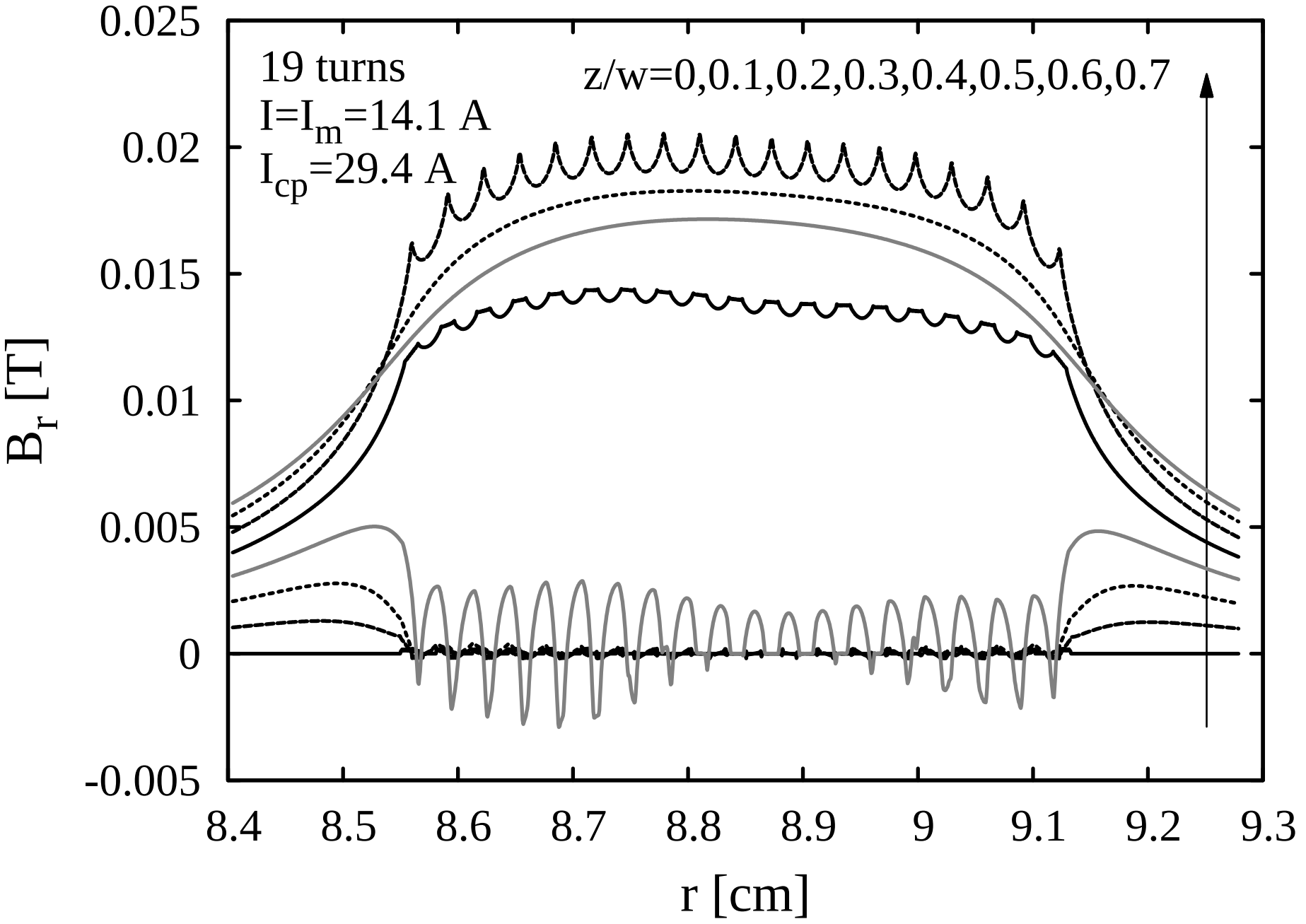

**Figure 4**

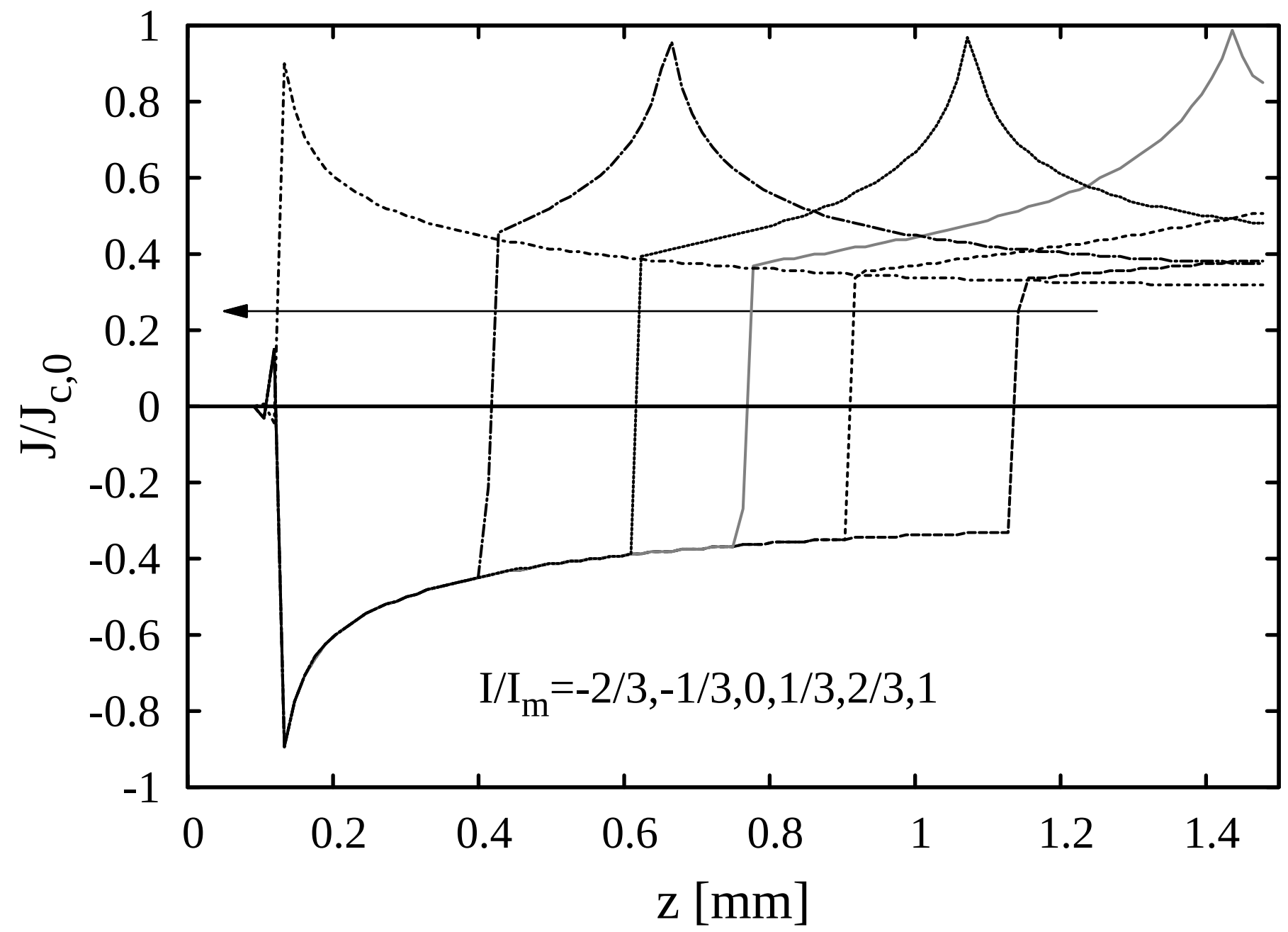

**Figure 5**

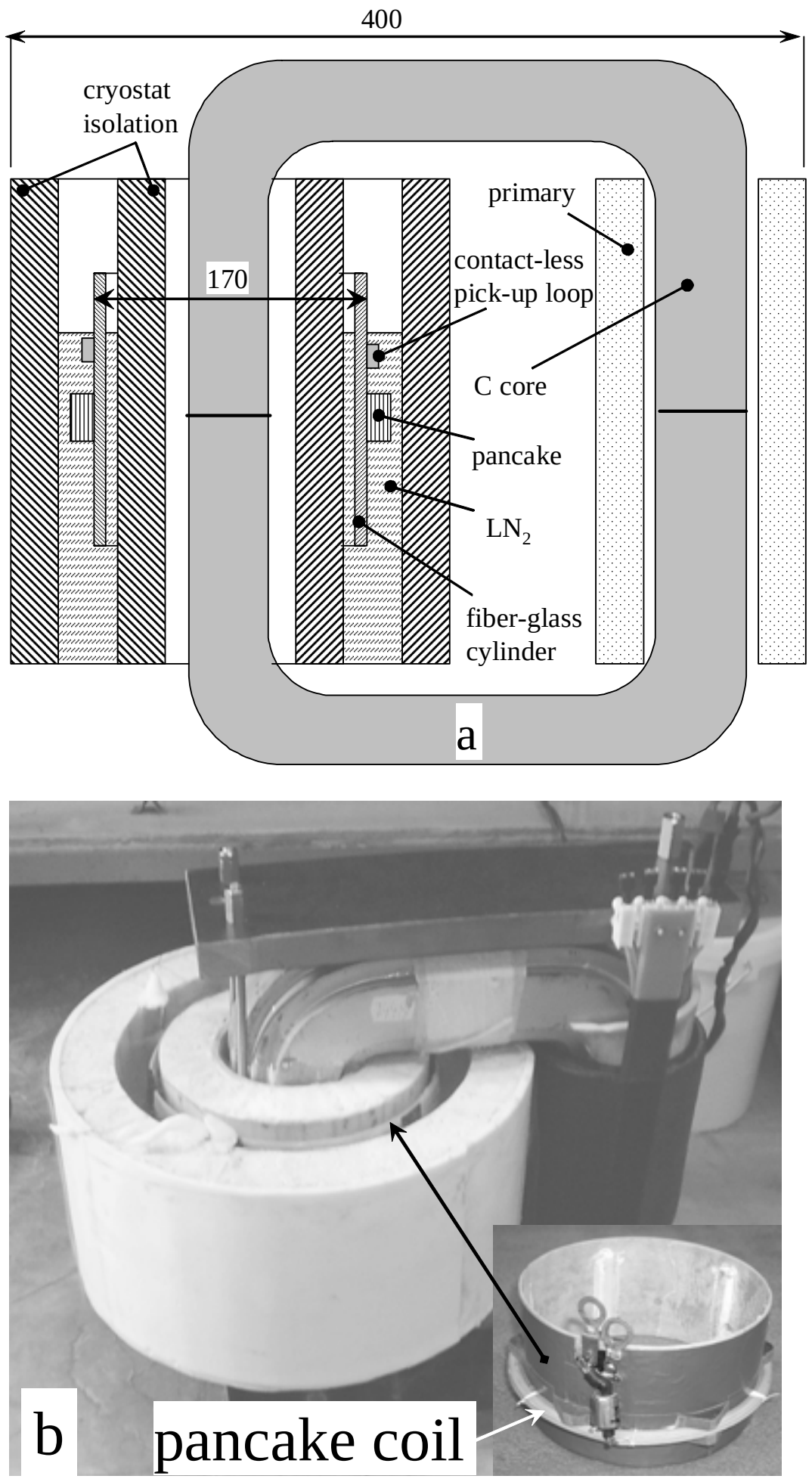

**Figure 6**

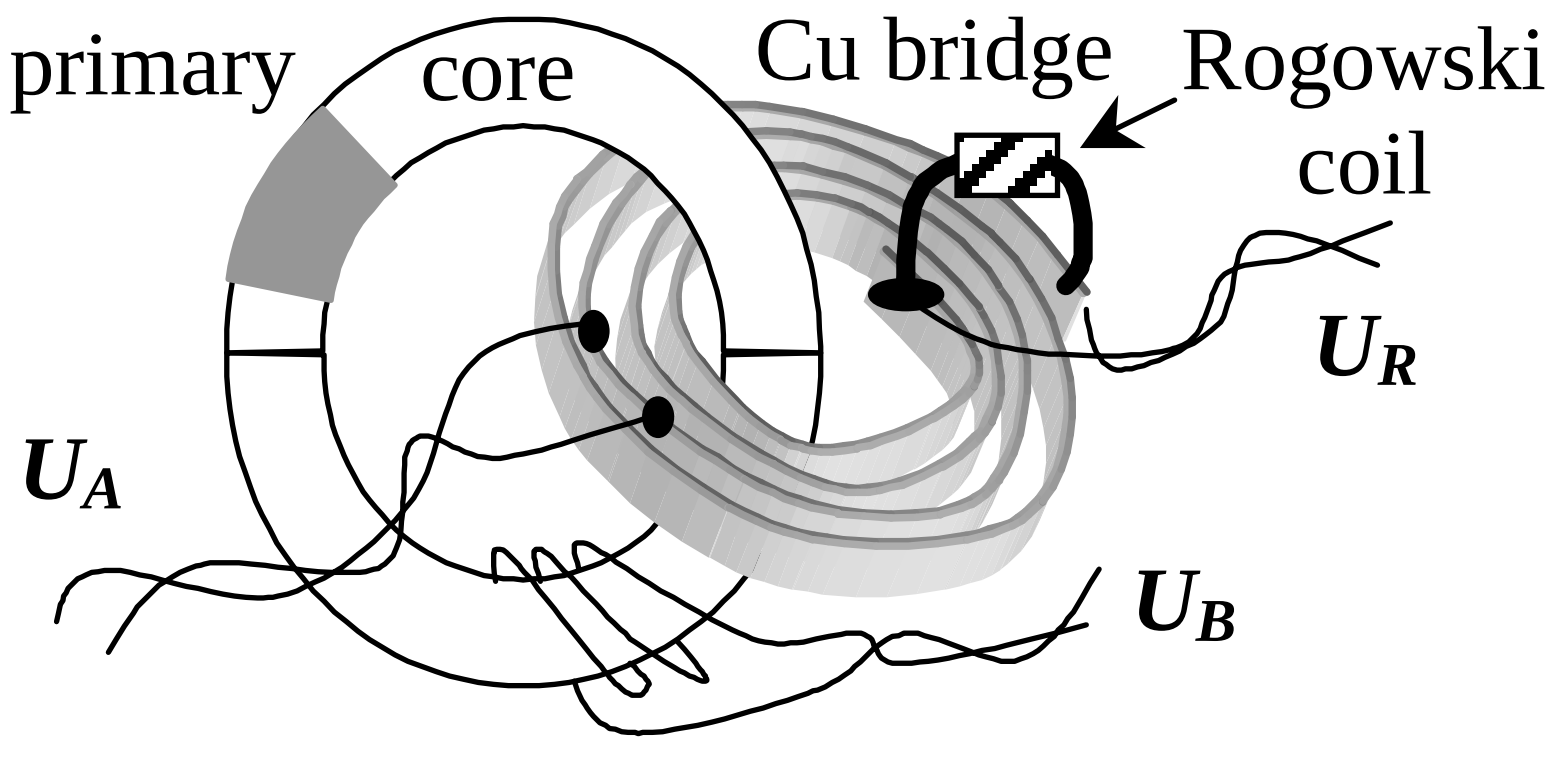

**Figure 7**

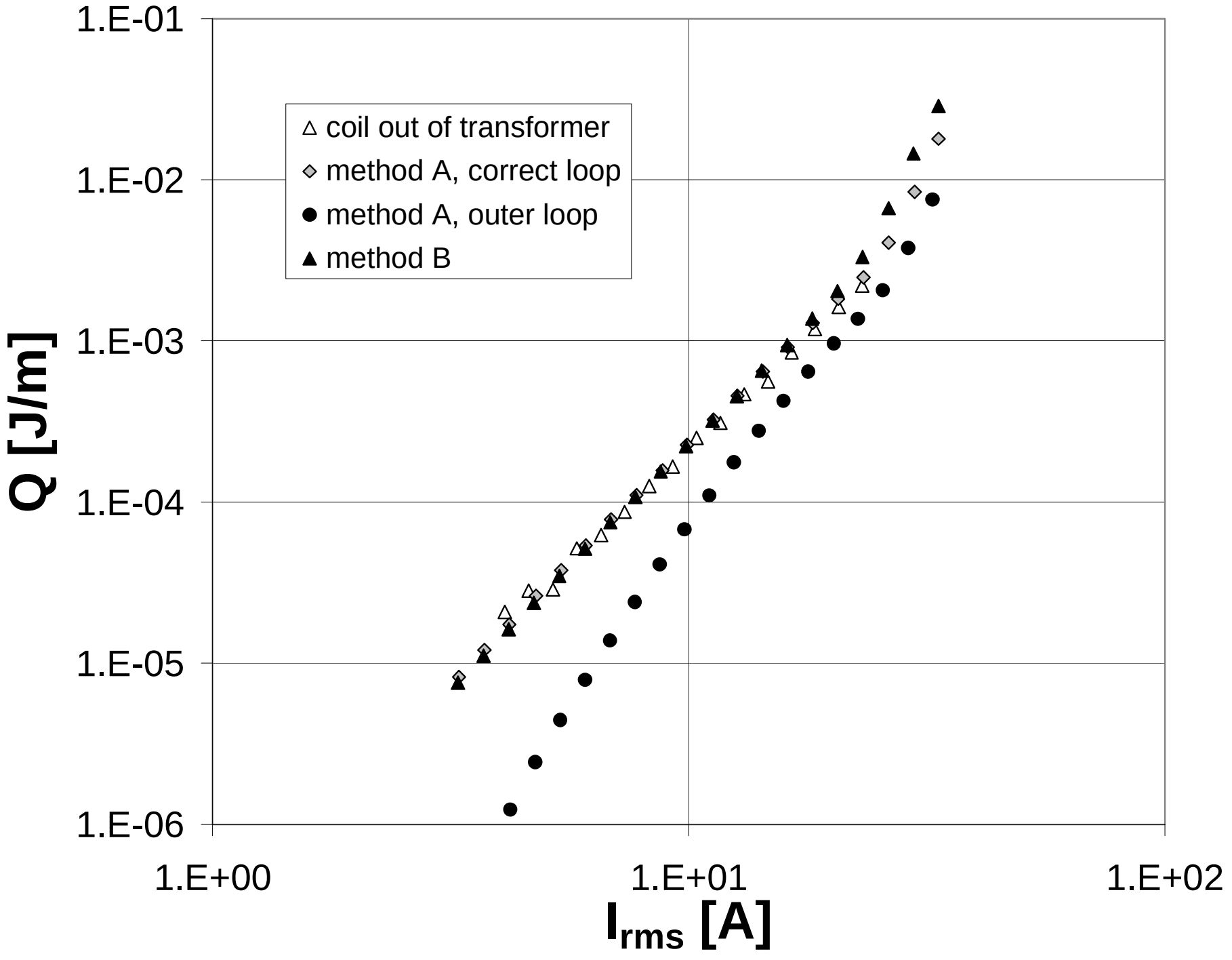

**Figure 8**

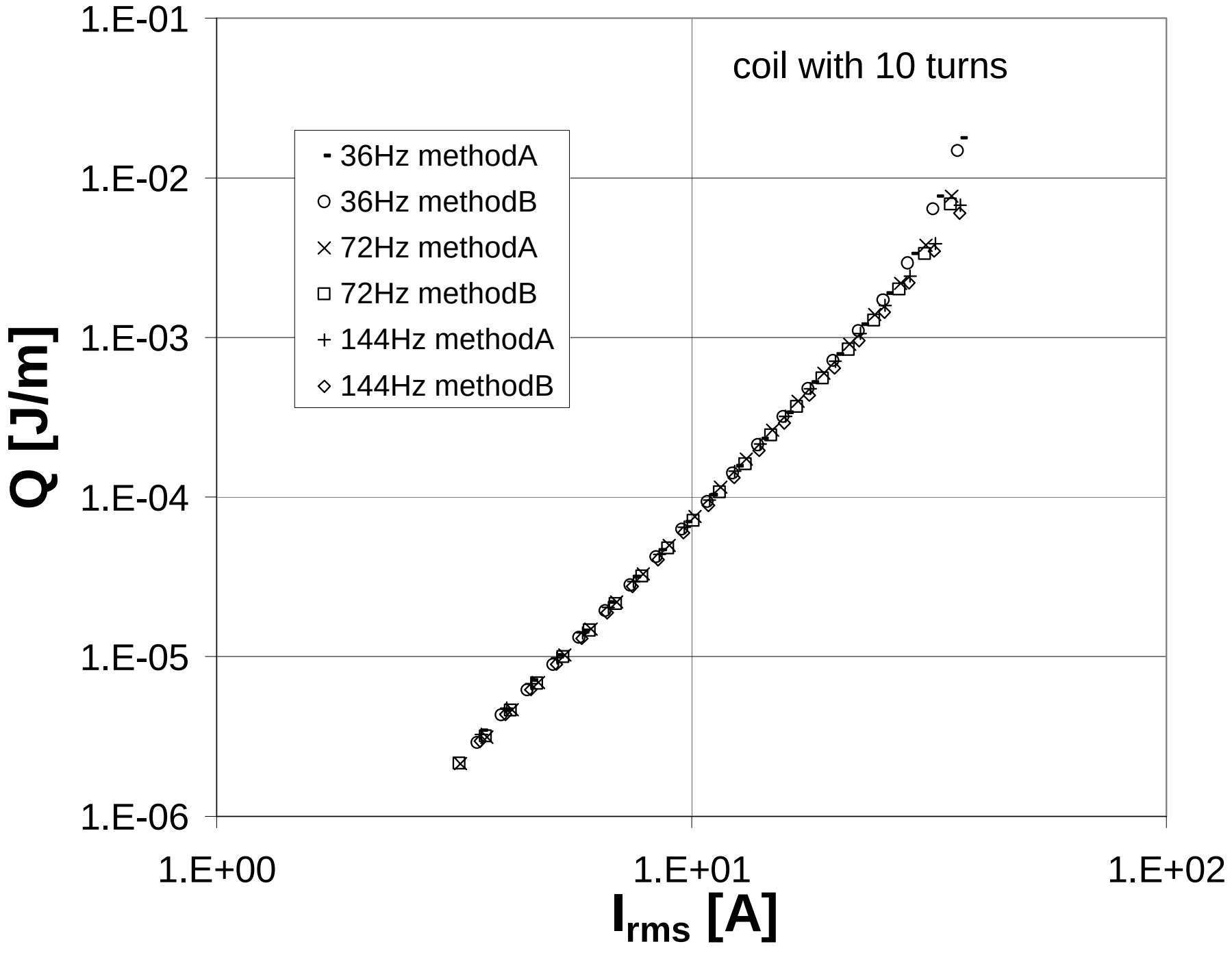

**Figure 9**

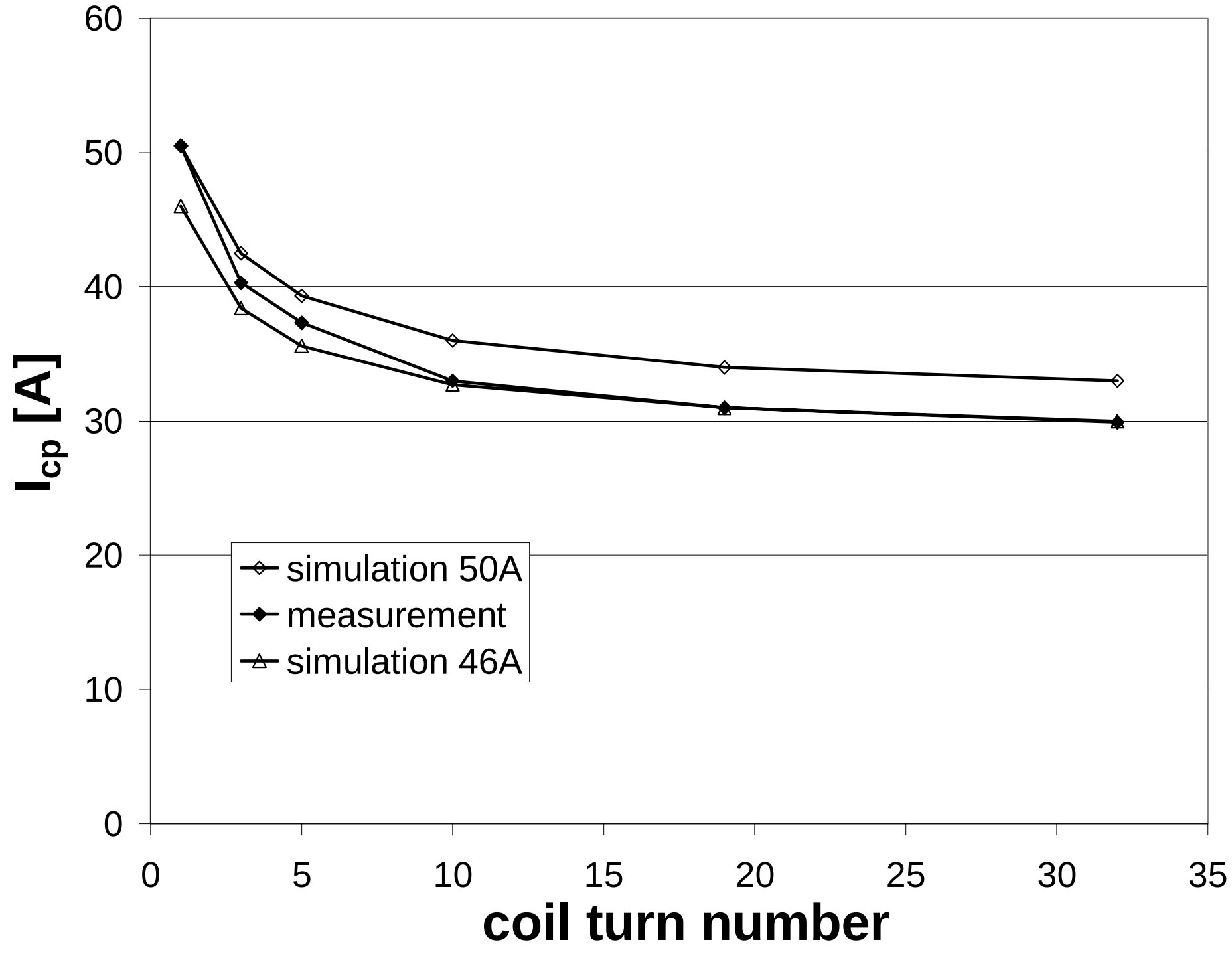

**Figure 10**

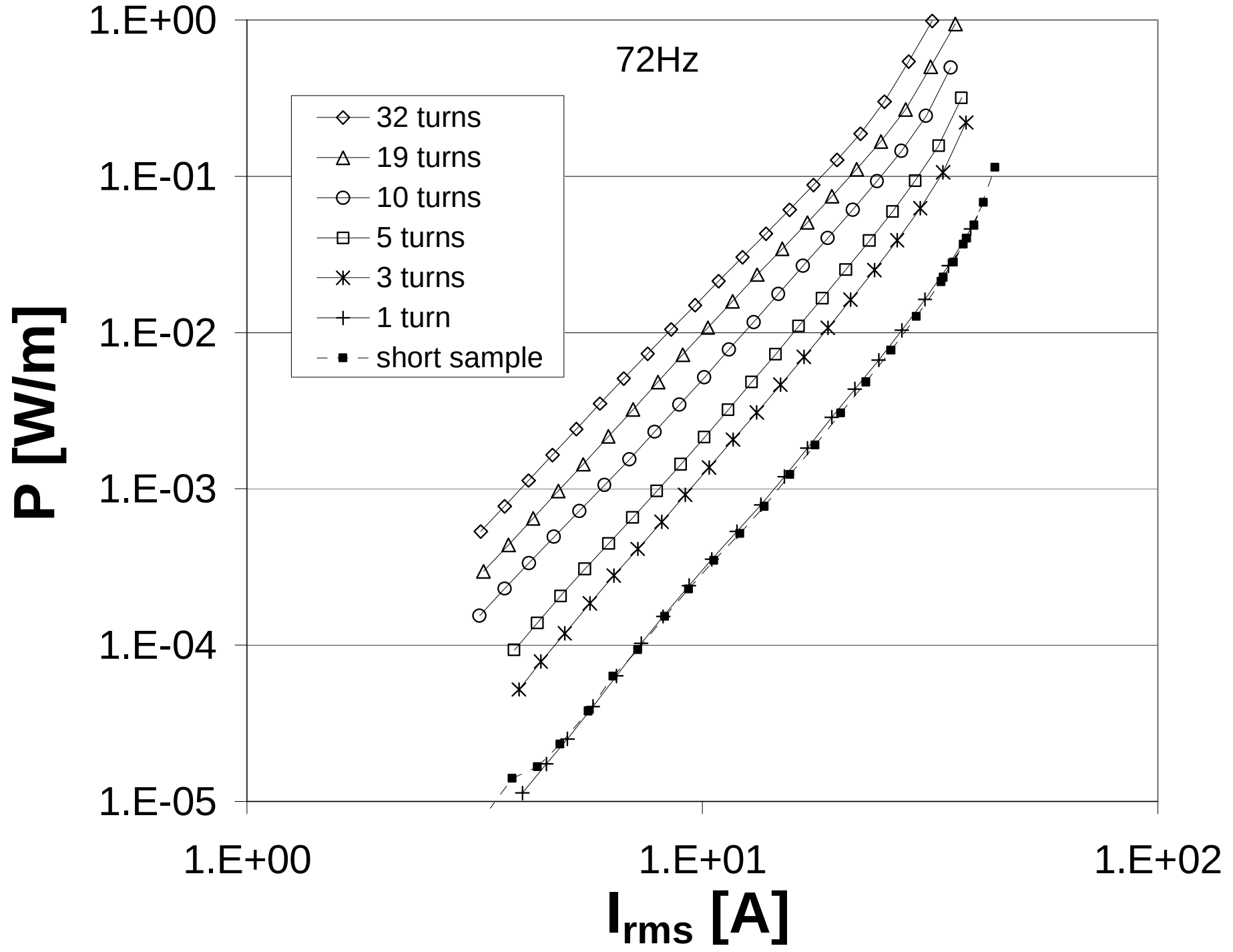

**Figure 11**

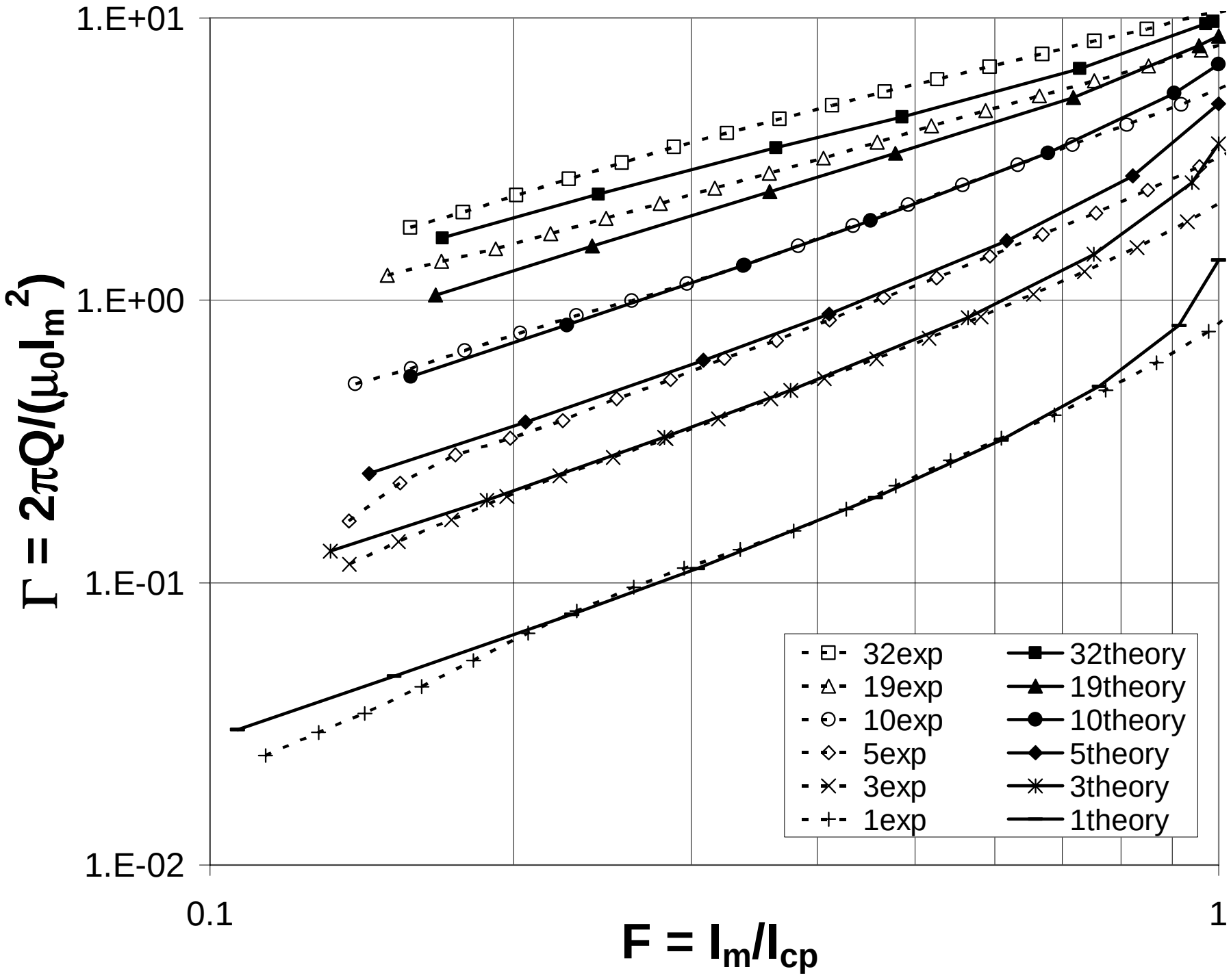

**Figure 12**

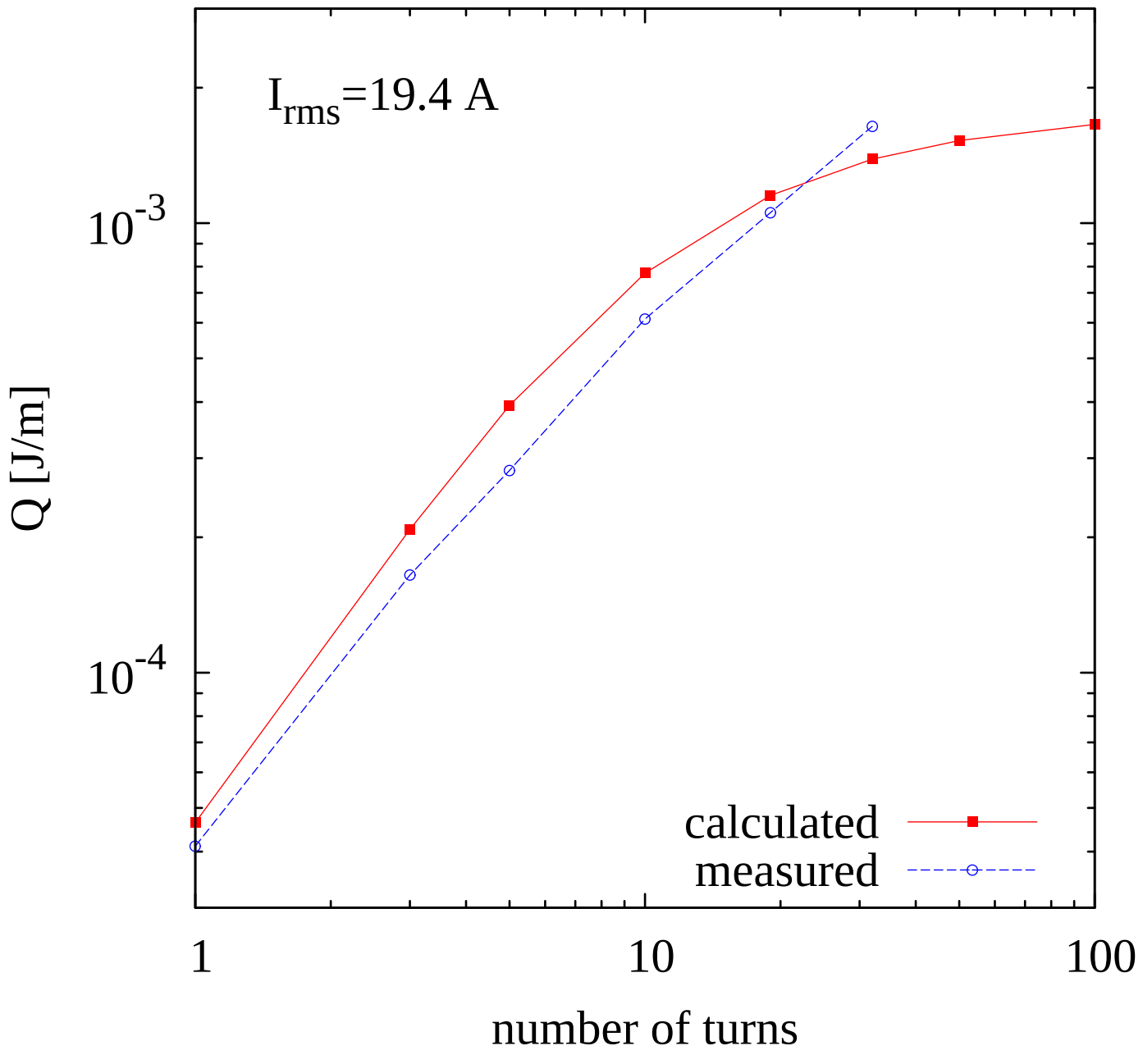

**Figure 13**

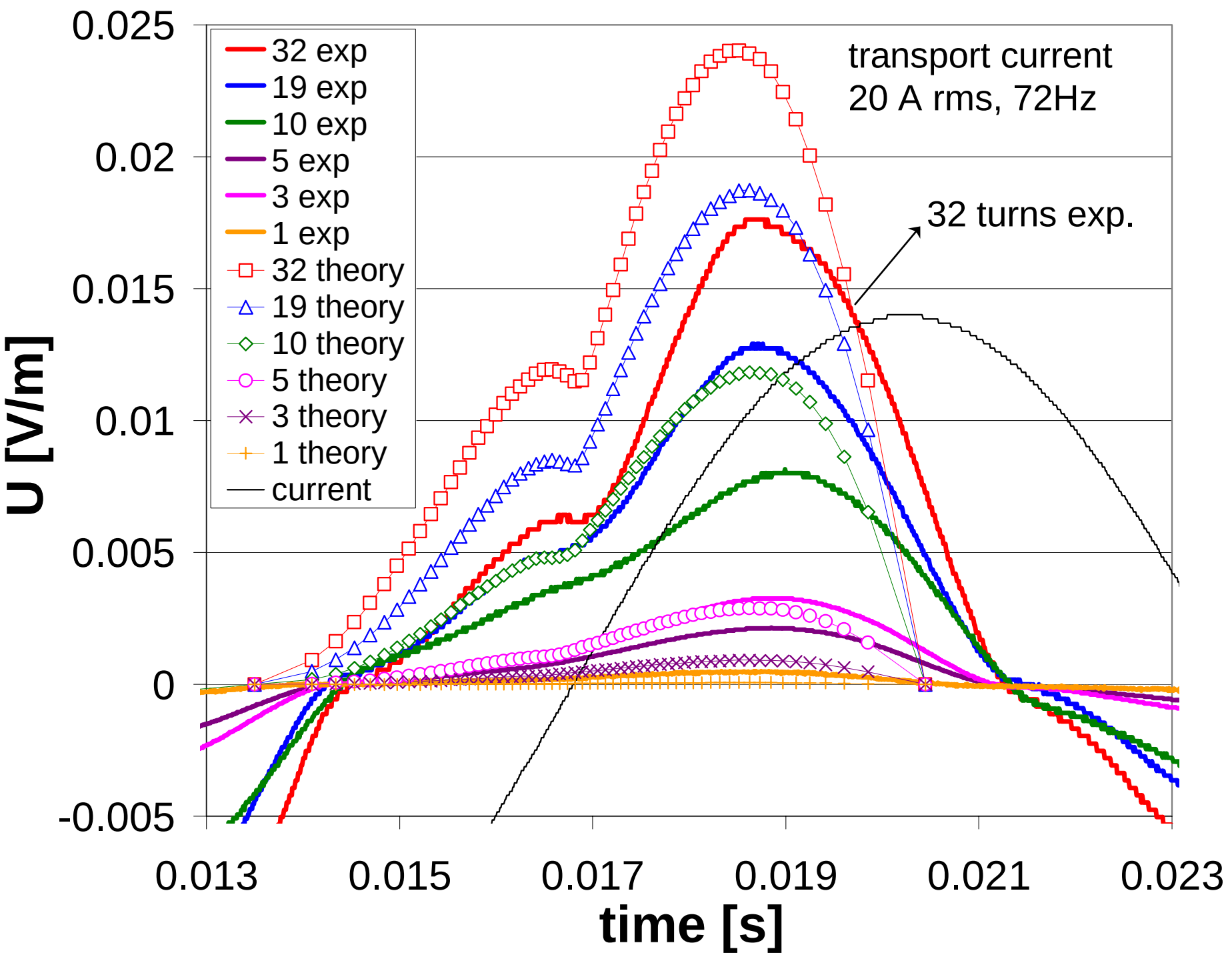

Figure 14

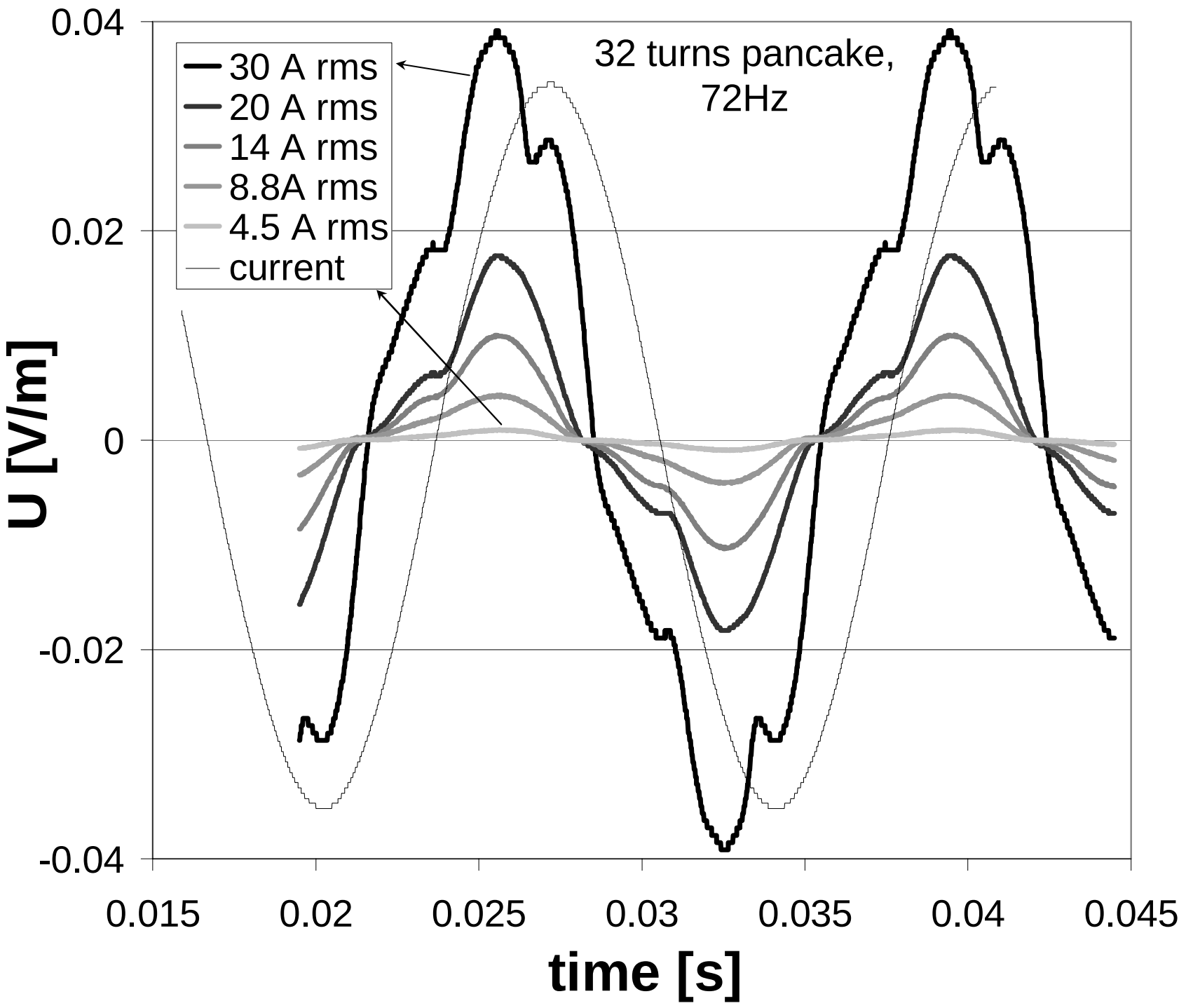

Figure 15